\documentclass[manuscript]{aastex}
\usepackage{emulateapj5}

\shortauthors{von Braun and Mateo}
\shorttitle{Variable Stars in NGC 3201}

\begin{document}

\title{Variable Stars in the Field of the Globular Cluster NGC 3201}

\author{Kaspar von Braun}
\affil{University of Michigan}
\affil{Department of Astronomy}
\affil{500 Church}
\affil{Ann Arbor, MI 48109-1090}
\email{kaspar@astro.lsa.umich.edu}

\and{}

\author{Mario Mateo}
\affil{University of Michigan}
\affil{Department of Astronomy}
\affil{500 Church}
\affil{Ann Arbor, MI 48109-1090}
\email{mateo@astro.lsa.umich.edu}

\begin{abstract}
We report on the discovery and analysis of 14 short-period variable
stars in the field of the southern globular cluster NGC 3201, located
within roughly two magnitudes on either side of the main-sequence
turnoff. 11 of these variable stars are eclipsing binaries, one is
an RR Lyrae, and two are thus far unclassified systems. Among the
eclipsing binary stars, nine are of the W Ursa Majoris (W UMa) type,
one an Algol (EA) system, and one a detached system. Using spectroscopic
follow-up observations as well as analysis of the variables' locations
in the color-magnitude diagram of the cluster, we find that only one
variable star (a W UMa type blue straggler) is actually a member of
NGC 3201. We present the phased photometry lightcurves for all the
variable star systems as well as their locations in the field-of-view
and in the color-magnitude diagram. 
\end{abstract}

\keywords{Globular Clusters: individual (NGC 3201); color-magnitude diagrams;
dust, extinction; binaries: eclipsing; blue stragglers; stars: variable:
general.}

\section{Introduction}

Variable stars have historically served as important tools and
{}``laboratories{}'' in our understanding of star formation, the
formation of stellar clusters, the calibration of distance
determination methods, and a variety of other areas. In particular,
the study of eclipsing binary stars (EBs) in a globular clusters (GC)
can provide a method of examining certain aspects of the GC itself,
and it may be used to obtain a value for the cluster's distance and a
constraint concerning turnoff masses of the GC stars (see, e.g.,
\citet{paczynski96}).

Simply detecting EBs in the fields of GCs and confirming cluster
membership is a straightforward - though data-intensive - task. These
systems expand the relatively meager sample of EBs which are currently
confirmed GC members (see for example
\citet{mateo1996,mcvean97,rubenstein97,rucinski2000}, and references
therein). A statistical evaluation of the number of known member EBs
in GCs can help in the determination of physical quantities such as
the binary frequency in GCs as a parameter in the study of dynamical
evolution of GCs (see for example \citet{hut92}), or in the
calibration of methods such as the Rucinski magnitudes of and
corresponding distances to W UMa binaries (see Section 4.3, and
\citet{rucinski1994,rucinski1995,rucinski2000}).

Moreover, the detection of blue straggler (BS) binary systems would assist
in shedding light on the binary frequency among this subclass of
stars, which, in turn, could help understand their formation and
evolution.  Some hypotheses in the study of the formation of blue
straggles claim that they have formed by collisions between single
stars, coalescence of binary systems, and/or interactions between
binary systems. Getting a handle on the frequency of binary blue
stragglers in a certain GC would therefore present a tool for probing
into the dynamical past and even the star formation history of the GC
itself (see for example \citet{leonard92,hut93,livio93,sills99,sills00}).

The simultaneous analysis of photometric and spectroscopic data for
individual EB systems can provide a direct estimate of the distance
to the system (see for example \citet{andersen91} and
\citet{paczynski96}), and thus, if the EB is a GC member,
to the GC itself. The main sources of error in the distance determination
are a) the relation between surface brightness and effective temperature
of the binary and b) the precise determination of the interstellar
reddening along the line of sight to the EB. The method itself, however,
is free of intermediate calibration steps and can provide direct distances
out to tens of kpc. In turn, the knowledge of the distances to GCs
can then be used to calibrate a variety of other methods, such as
the relation between luminosity and metallicity for RR Lyrae stars. 

The very same analysis can, in principle, be used to obtain the
Population II masses of the individual components of the EB system
(see Paczynski 1996). If mass is conserved, the evolution of any EB
may be expressed by \( M_{1}+M_{2}=M_{1,0}+M_{2,0}+M_{L} \), where \(
M_{i} \) are the present day masses of the two components, \( M_{i,0}
\) are their initial masses, and \( M_{L} \) is the total mass loss
from the system.  If one assumes that \( M_{L}=0 \) (no mass loss),
then the difference between present day masses and initial masses of
the components is only due to the mass transfer history between the
two stars. This demostrates the value of detecting \textit{unevolved}
EB systems.  These are systems where the binary components are
detached, no mass transfer has taken place, and the present day masses
are therefore equal to the initial masses.

A direct determination of the ages of the systems, however, remains a
challenging task. The reasons for this are described in
\citet{paczynski96}; we will briefly repeat them here. Analysis of
various theoretical isochrones (e.g., \citet{bertelli94},
D. VandenBerg 2000, private communication; based on evolutionary
models by \citet{vandenberg00}) readily shows that at the
main-sequence turnoff,

\begin{equation}
age_{MSTO}\sim Lum^{-1}_{MSTO}\sim Mass^{-3.7}_{MSTO}.
\end{equation}  

From Kepler's third law, it is apparent that \( M_{1}\sim K_{2}^{3}
\), where \( M_{1} \) is the mass of the first component and \( K_{2}
\) is the radial velocity amplitude of the second component. Combining
these two relations gives \( age_{MSTO}\sim K^{11} \), implying a 12\%
uncertainty in age for a 1\% uncertainty in the velocity amplitude!  A
perhaps more rewarding approach would therefore be to use the MSTO
masses to provide a fundamental check of stellar models at low
{[}Fe/H{]}.  In order to do this, one needs to either detect detached
EB systems (and assume no mass loss, in which case the zero-age masses
equal the MSTO masses of the two components), or reconstruct the
history of mass transfer between contact EBs to obtain zero-age masses
for the two stars.

We are currently undertaking a survey of approximately 10 Galactic GCs
with the aim of identifying photometrically variable EBs around or
below the MSTO. Our observing strategy, aimed at detecting binaries in
the period range of approximately 0.2 to 5 days (see \citet{hut92})
basically consists of repeated observations of a set of GCs during
each night. Multiple observing runs are usually helpful in detecting
variables with a period of close to one day (or to a multiple
thereof). The more valuable detached systems are generally much harder
to detect because a) they will eclipse each other only within a small
range of inclination angles due to the larger distances between the
components, b) their duty cycle\footnote{ The duty cycle of an EB is
defined as the fraction of the orbital period during which the system
is experiencing an eclipse for an edge-on system (inclination angle \(
i=90^{\circ } \)).} is very low, and c) they vary less in brightness
(if at all) inbetween eclipses than the more distorted components of
contact systems. Fairly extensive coverage is therefore usually needed
in order to detect detached EB systems.

NGC 3201, by chance one of the first GCs in our sample that we
analyzed, has been probed for the existence of binary stars in the
past, most recently by \citet{cote1994}. The results of even earlier
work on variables in NGC 3201 are summarized in \citet{sawyer-hogg73}
and  \citet{clement01}, and references therein. The
magnitude and period ranges of the EBs covered in these earlier
studies, however, does not overlap with the work presented here.

Details on our photometry observations are given in Section 2. We then
describe how we detect variable stars in our sample and determine
their periods in Section 3. These two Sections set out the methods we
will use for the other clusters in our sample. Section 4.1 contains
our results concerning the locations of the binaries in the field as
well as in the CMD of NGC 3201. In Section 4.2, we present the phased
lightcurves for all the variable stars we find. We calculate distance
moduli to our set of W UMa type EBs using Rucinski's formula
\citep{rucinski1994,rucinski1995,rucinski2000} in Section 4.3. Section
4.4 reports our spectroscopy results concerning the membership of our
variable stars. We finally summarize and conclude with Section 5.

\section{Observations and Photometry Reductions}

\subsection{Initial Photometry Observations to Find Variables}

\begin{figure}
{\centering \includegraphics{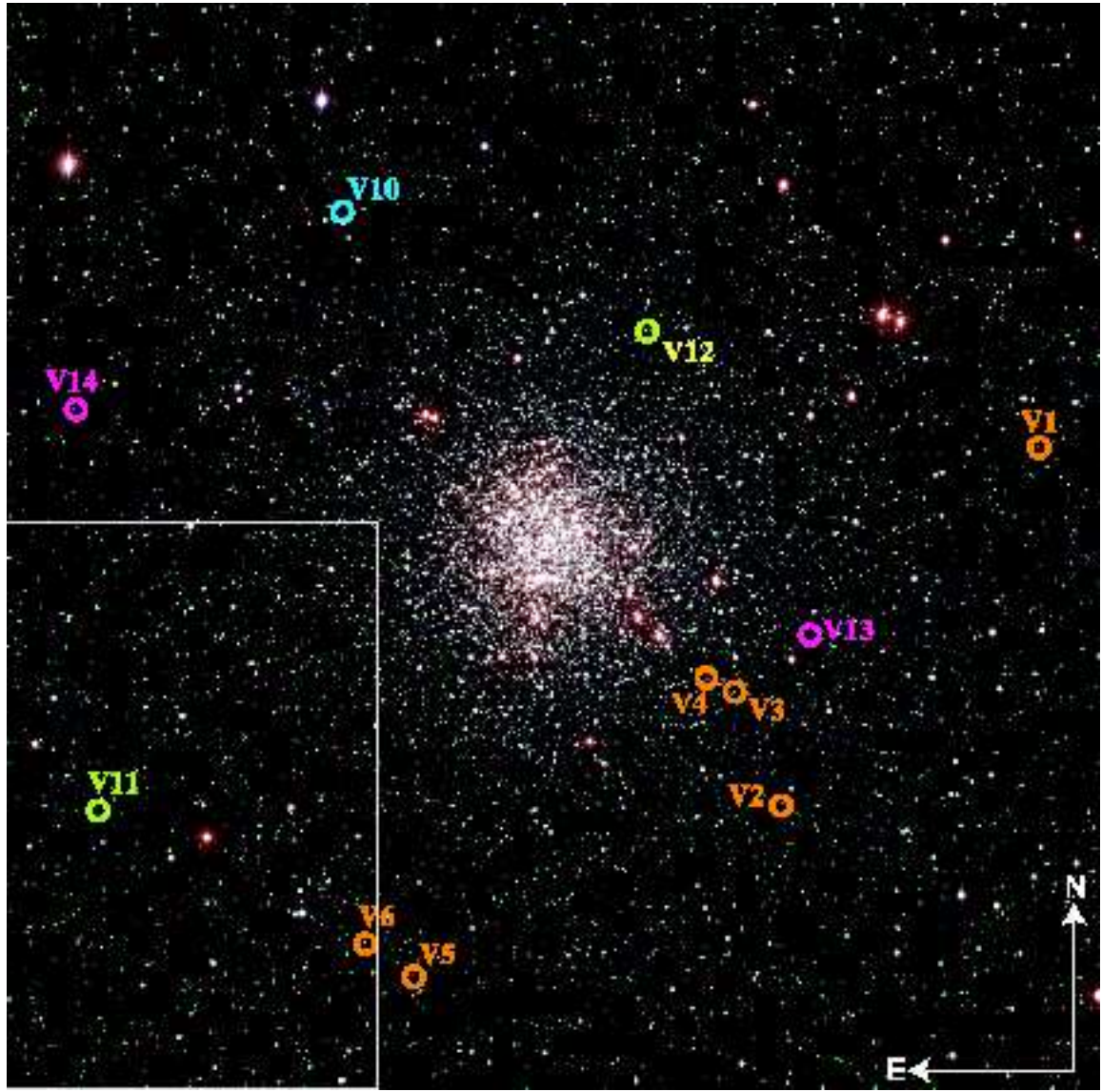} \par}
\caption{\label{vars_location}The locations of the variable stars in
the LCO data. The field-of-view is 23.5 arcmin on the side. North is
up and east is to the left. The rectangle toward the bottom left part
of the figure represents the the approximate boundaries of the overlap
region of this Figure and Figure \ref{fieldA}, i.e., the north-west part 
of Figure \ref{fieldA}.}
\end{figure}

The bulk of the photometry observations for NGC 3201 was conducted
during three separate observing runs in June 1996, May 1997, and May
1998 at the Las Campanas Observatory (LCO) 1 m Swope Telescope. For
all three runs Johnson-Cousins $VI$ filters were used, and
the number of images are approximately evenly divided between the
two bands. During the June 1996 run we obtained 17 epochs (600s exposure
time) using a TEK 5 CCD with 24 arcmin on the side. For both the May
1997 and May 1998 runs we used a SITe 1 CCD with 23.5 arcmin on a
side (Figure \ref{vars_location}), and obtained 164 epochs and 89
epochs (all with 600s exposure time), respectively. During the May
1998 run we took additional shorter exposures of the cluster to complete
the CMD in the brighter regions (these shorter exposures were not
used to detect variable candidates), as described 
in \citet[BM01 hereafter]{BM01}.

\subsection{Processing and Data Reduction of the LCO Data}

The details of the IRAF\footnote{IRAF is distributed by the National
Optical Astronomy Observatories, which are operated by the Association
of Universities for Research in Astronomy, Inc, under cooperative
agreement with the NSF.} processing as well as the basic data
reduction of the May 1998 data are described in BM01, section 2.1. The
processing of the May 1997 and June 1996 runs was performed in the
same way. All data were reduced using DoPHOT \citep{schechter93} in
fixed position mode (see BM01) after aligning every image to a
deep-photometry template image created by combining the 15 best seeing
frames for each filter from the May 1998 data. The astrometric
measurements, application of aperture corrections, as well as
photometric calibration of the May 1998 images are outlined in
BM01. All data from the other two observing runs were shifted to the
coordinates and the photometric system of the calibrated May 1998 600s
exposures. Since NGC 3201 suffers significant differential reddening
across the field-of-view, the stars were dereddened using the
differential extinction map in BM01 so that the $E_{V-I}$ zero point
applied to all stars in the field is 0.15.  This differential
reddening correction is applied to the data presented in Figure
\ref{cmd_vars}.

\subsection{Additional Photometry}

\begin{figure}
{\centering
\resizebox*{12cm}{12cm}{\includegraphics{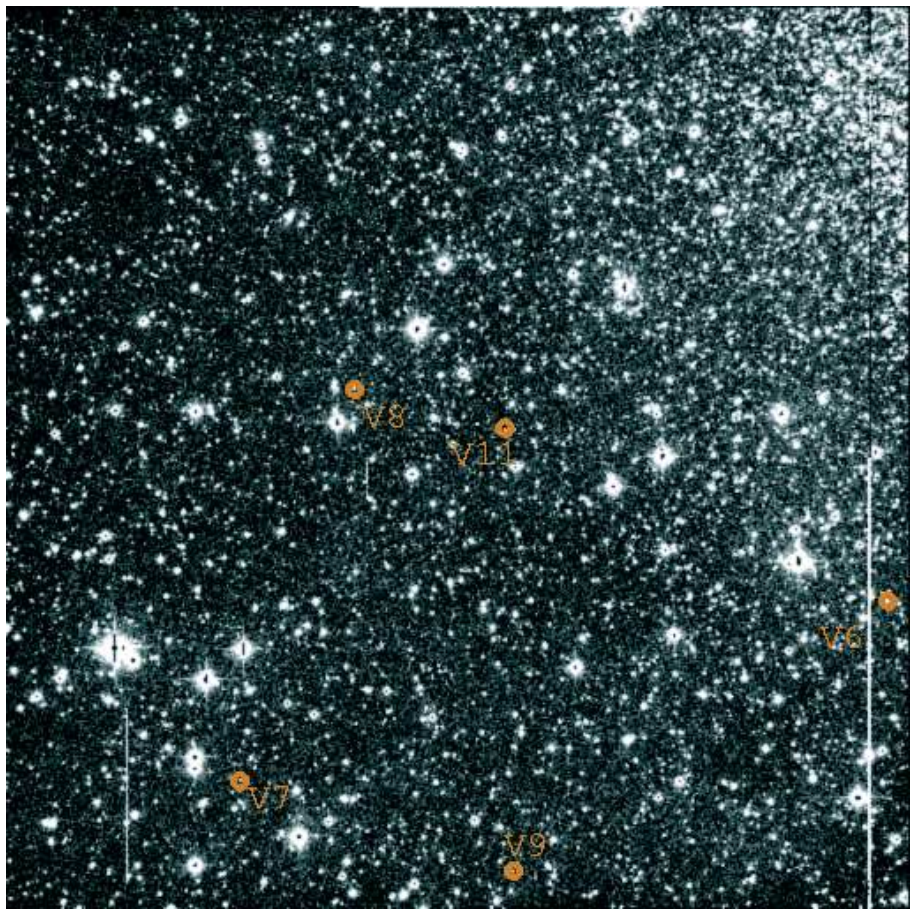}} \par}
\caption{\label{fieldA} The locations of the variables extracted from
the CTIO 0.9m data. The field-of-view has the same orientation as the
one in Figure \ref{vars_location}, but is smaller (13.5 arcmin on the
side in this figure). V11, toward the center of this image, and V6,
toward the right edge of the image, are present in both
fields-of-view. V7-9 are located outside (south-east of) the field of
Figure \ref{vars_location}.}
\end{figure}

\begin{figure}
{\centering \resizebox*{0.6\textwidth}{0.4\textheight}{\includegraphics{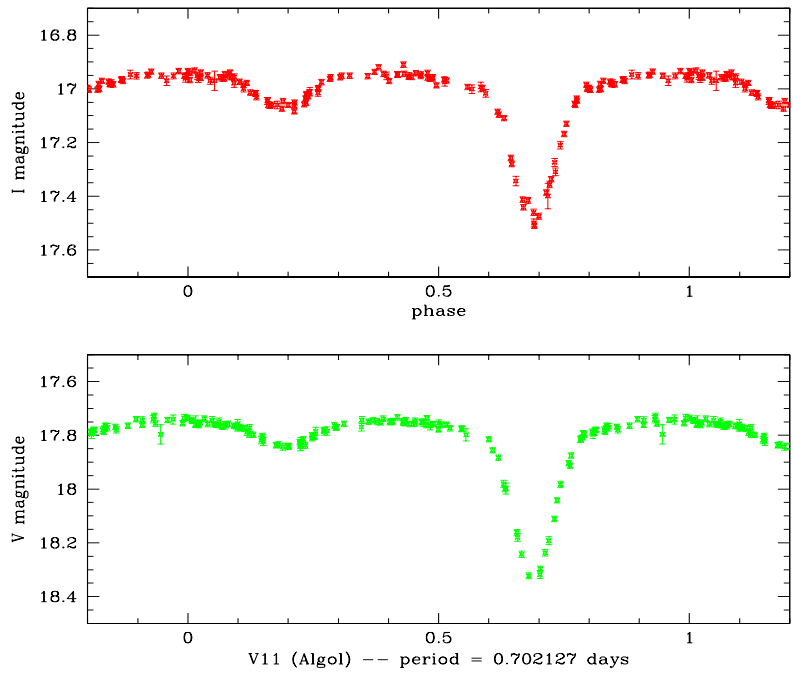}} \par}
\caption{\label{algol}The lightcurve of V11, the Algol system,
comprised of data taken at both LCO and CTIO. No reddening correction
is applied to the data. V11 is not a member of the globular cluster. }
\end{figure}

In order to precisely determine the times of quadrature (maximum
radial velocity) for two of the variable stars we discovered (V11, the
Algol, and V12, the detached system; see Section 4.1 and Figures
\ref{algol} and \ref{detached}), we performed additional $VI$ 600s
photometry observations at the Cerro Tololo Inter-American Observatory
(CTIO) 0.9m Telescope during the nights of February 25 - 28, 2001,
just prior to our spectroscopy follow-up CTIO 4m run (see Section
4.4). The 0.9m telescope setup included a TEK2048 CCD which provided a
field-of-view of approximately 13.5 arcmin on the side (Figure
\ref{fieldA}). Initial data processing was done with the IRAF QUADPROC
package. For each night, 10 bias frames were combined for the bias
subtraction. Flat-field images were produced by combining five
twilight flats per filter per night. The fields we observed were
centered on the approximate locations of the detached candidate (Field
D: \( \alpha _{2000}=10^{h}17^{m}25.72^{s} \) and \( \delta
_{2000}=-46^{\circ }20^{'}12.4^{''} \)) and the Algol candidate (Field
A: \( \alpha _{2000}=10^{h}18^{m}32.74^{s} \) and \( \delta
_{2000}=-46^{\circ }30^{'}42.5^{''} \)). Field D fell entirely within
our original field-of-view which was centered at \( \alpha
_{2000}=10^{h}17^{m}36.8^{s} \) and \( \delta _{2000}=-46^{\circ
}20^{'}40.0^{''} \), whereas Field A contained stars which we had not
observed previously (see Fig. \ref{fieldA}).  We secured 31 $I$ and 27
$V$ observations of Field A as well as 33 $I$ and 35 $V$ observations
of Field D.

Following the publication of this work, we will make our entire 
photometry dataset available to readers over the Internet, using 
NASA's Astronomical Data Center, accessible at 
\url{http://adc.gsfc.nasa.gov}.

\section{Finding the Variable Stars}

The starting point for our analysis of the LCO photometry with respect
to finding binaries and determining their periods was a database which
contained the data on approximately 130 observational epochs (600s
exposure time) of 45000 stars in each filter per image.

\subsection{Variability Detection}

The criteria we set in order to extract variable star candidates from
the list of stars in our database were the following:

\begin{enumerate}
\item $\chi^{2}$ per degree of freedom, calculated based on the
assumption that every star is a non-variable, has to be greater than
3.0. We furthermore set a \( \sigma \) $>$ 0.05 mag threshold for a
star to be taken into consideration as a variable candidate.
\item The star under investigation must appear in at least 75\% of the epochs. 
\item The detected variability should not be due to only a few outliers
(3-5\% of the datapoints) causing the high $\chi^{2}$. Care was taken
to avoid deleting possible eclipsing systems at this stage (see below).
\item The star under investigation should display a brightness variation
in both filters, and the variability signal should be correlated in
both filters (this algorithm is very similar to the one described
in \citet{WS93} and \citet{S96}).
\end{enumerate}

Any measurement of stellar magnitude was weighted by the square of the
inverse photometric error associated with it. Steps 1 and 2, which
were performed simultaneously, returned around 2700 ($V$) and 4500
($I$) variable candidates. Step 3 reduced both of these numbers by
approximately 50\%. Step 4 further reduced the number of candidates to
a total of 80 candidates. The data for these remaining candidates were
then phased and inspected by eye, which is described in the next
Subsection.

We note that we did not systematically address the issue of
completeness in our analysis. The choice of parameters for the steps
outlined above was made mainly based on hindsight (e.g., all of our
candidates' $\chi^{2}$ values are well in excess of 100) or trial and
error (e.g., the analysis of a set of phased lightcurves before and
after a reduction criterion was applied).

One set of binary star system, however, might be prone to being
deleted during step 3. A detached system with a period long enough so
that eclipses are not very well sampled might show up as a system
whose $\chi^{2}$ is based solely on a few outliers. In order to ensure
that we would not miss such a system, we developed a slightly modified
set of criteria specifically aimed at detecting detached binary star
systems with the basic idea to identify faint {}``outliers{}'' which
fall next to each other in time.

\begin{enumerate}
\item As a first step, the mean magnitude plus standard deviation of the
brightest 75\% of the datapoints are calculated.
\item We now try to detect successive (in time) datapoints which are
n\( \sigma \) fainter than this mean magnitude. The choice of n
depends on the quality of the photometry of the star in question and
the number of datapoints for the star.
\item If the $i$-th datapoint falls within \( \Delta t \) of datapoint
$i-1$ (where $i$ indicates succession in time), a merit function is
calculated whose value would be $\delta_{i}\times \delta_{i-1}$.
$\delta_{i}$ is defined as the magnitude difference between the $i$-th
datapoint and the average magnitude of the 75\% of the brightest
stars, divided by the standard deviation (see step 1). \( \Delta t \)
is dependent on the observing strategy followed to detect the binaries
(i.e., it would be shorter, perhaps up to an hour, for a strategy
where the cluster is observed many times in succession, and longer,
perhaps several hours, if the cluster is observed only once every so
often). It is obvious that for more than two neighboring faint
outliers, the merit function would rapidly increase in value.
\item Candidates for detached systems are selected based on the value of
their merit functions.
\end{enumerate}

The application of this algorithm to our data revealed the existence
of one detached system in our field-of-view (V11; see Section 4.1 and
Figure \ref{detached}) which had been eliminated by application of
criteria described at the beginning of this Subsection.

\subsection{Period Determination}

The final decision as to whether a star is a true variable candidate
(and if so, what kind of variable star it is) was based on the
inspection of the phased data (the photometry lightcurve). The initial
estimates of the periods of all our variable star candidates which
survived the steps outlined in the previous Subsection were determined
by two independent algorithms:

\begin{itemize}
\item the minimum-string-length (MSL) method, based on a technique by
\citet{LK65} and described in \citet{S96}.
\item the Analysis of Variance (AoV) method, described, e.g., in
\citet{SC89}. The basic code for this algorithm was supplied to us by
Andrzej Udalski (private communication, 1998).
\end{itemize}

Both of these methods worked very well with our dataset to the extent
that the periods for a variable star candidate, as independently
determined by the two methods, were practically identical. A final
tweaking of the precision in the period for a candidate was then done
by hand, based on the appearance of the lightcurve. In the majority of
the cases of the smoothly varying variables, such as W UMa stars or
the RR Lyrae, the periods as determined by AoV or MSL could, in
principle, not be improved by hand, i.e., the period was already
determined to the maximum precision possible for our dataset. In the
cases of the detached system and the Algol, we were able to increase
the precision of the periods in general at the $\geq 10^{-4}$ days
level. This effort helped ensure that knew the exact times of
quadrature of these systems during our spectroscopic observations.

\subsection{Analysis of Additional CTIO Photometry Data}

In order to verify that our period estimates for the detached (V12)
and Algol (V11) systems were sufficiently precise to accordingly phase
our spectroscopic follow-up observations, we re-observed NGC 3201 at
the CTIO 0.9m Telescope, as described in Subsection 2.3. Field A and
Field D provided additional phasing information for the two binaries,
enabling us to improve our period estimates for both of these
systems. In the case of the detached system, this correction was on
the order of five seconds, whereas the previously calculated period
for the Algol system proved to be off by less 0.2 seconds.

Field A furthermore contained stars we had not observed in our
previous monitoring program, giving us the chance to mine additional
data of NGC 3201 for the existence of variable stars. Using the
algorithms described above, we detected three additional W UMa systems
outside our previous field of view (V7-9; see Figures
\ref{vars_location} and \ref{fieldA}). We made use of the fact that we
had repeated $V$ and $I$ observations of the same system (the Algol
star) in both the CTIO and the LCO datasets to photometrically shift
the CTIO data to the system of the LCO data. The differential
dereddening of the CTIO dataset W UMa variables (V7-9) was performed
as described above and in BM01.

\section{Results}

\subsection{Locations of Variable Stars in Field and in CMD}

\begin{figure}
{\centering
\resizebox*{0.8\textwidth}{0.6\textheight}{\includegraphics{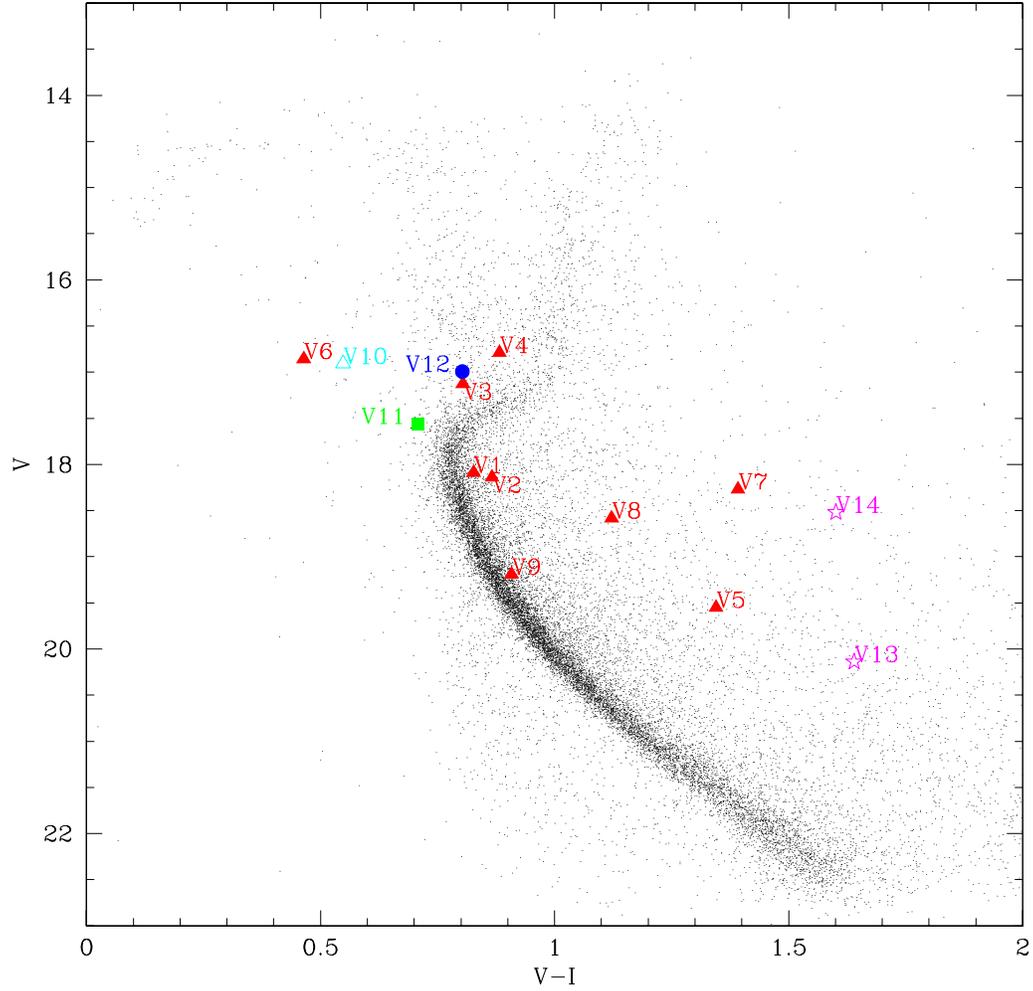}}
\par}
\caption{\label{cmd_vars} The locations of all our variable stars in
the field of NGC 3201 in its CMD. The data presented are
differentially dereddened to the fiducial region defined in BM01,
i.e., no reddening zero point is applied. In BM01, we calculate the
$E_{V-I}$ zero point to be 0.15. The variable stars are plotted at
maximum brightness and are also differentially dereddened. Filled
triangles represent W UMa types, the open triangle the RR Lyrae, the
square the Algol system, the filled circle the detached system, and
the open star-shaped symbols the unclassified systems.}
\end{figure}

Table 1 gives the basic information about the variable stars we
detected in the field of NGC 3201. $V_{bright}$ and $I_{bright}$ are
the $V$ and $I$ magnitudes at maximum light, respectively. Figure
\ref{vars_location} shows the locations of the variable stars in the
LCO dataset. Figure \ref{fieldA} shows the locations of V11 (the
Algol) and the three additional W UMa systems we found in the CTIO
photometry data. Finally, Figure \ref{cmd_vars} indicates where the
variables fall within the CMD of NGC 3201. The data shown in Figure
\ref{cmd_vars} are differentially dereddened to a fiducial region
within the cluster, as described in BM01. No reddening zero point is
applied.

\subsection{Photometry Lightcurves}

\begin{figure}
{\centering
\resizebox*{0.8\textwidth}{0.8\textheight}{\includegraphics{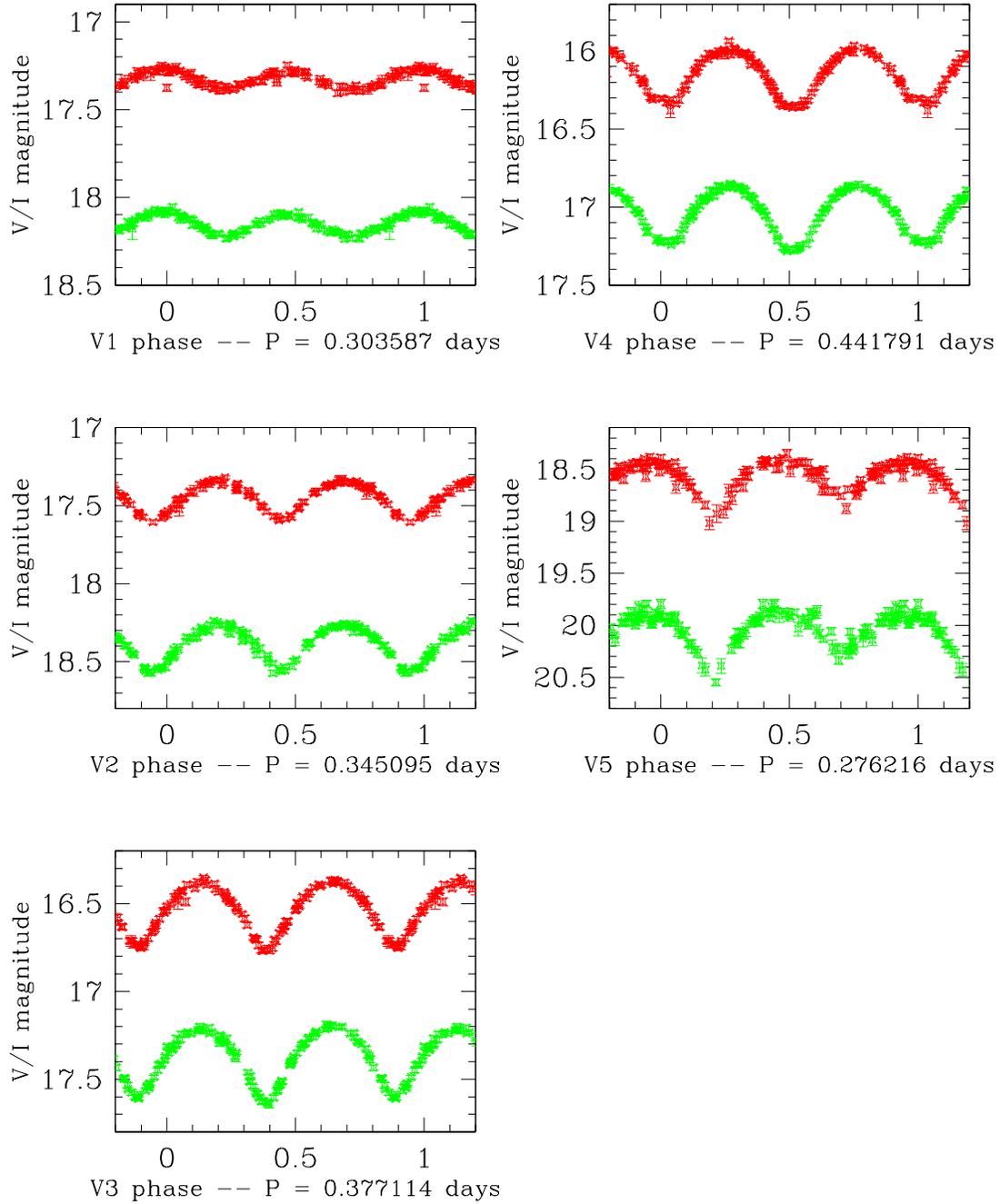}}
\par}
\caption{\label{lco_vars}The $V$ and $I$ lightcurves of V1 through V5
(all W UMa type binaries). $I$ magnitude is the upper curve in each of
the panels; no extinction correction is applied.  None of these
variables is associated with NGC 3201 itself. }
\end{figure}

\begin{figure}
{\centering
\resizebox*{0.8\textwidth}{0.8\textheight}{\includegraphics{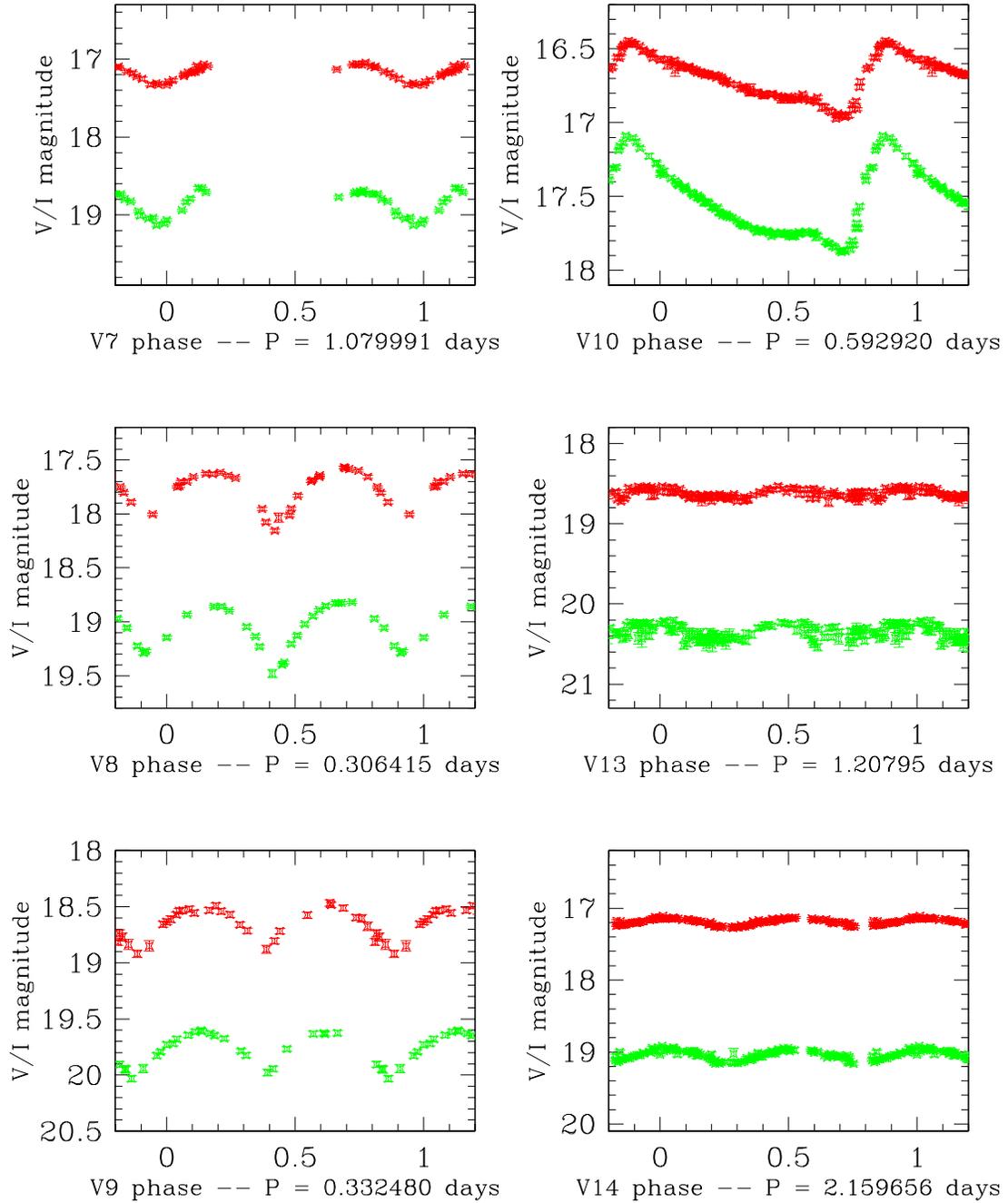}}
\par}
\caption{\label{var_rest}V7-V9 are the W UMa systems discovered in the
CTIO dataset. Note the much sparser sampling of the lightcurves. V10
is the background RR Lyrae star. Finally, V13 and V14 are the two
unclassified systems. Their lightcurves look similar to the W UMa
types, but their periods are much longer and their colors much
redder. In each panel, the $I$ magnitude is plotted above the $V$
magnitude.  The data are not corrected for extinction.}
\end{figure}

The phased lightcurves for the W UMa systems V1 through V5, detected
in the LCO dataset, are presented in Figure \ref{lco_vars}. The three
additional W UMa lightcurves, V7 through V9, extracted from the CTIO
photometry data (note the difference in sampling due to the lower
number of observational epochs for these stars), are in Figure
\ref{var_rest}, along with the lightcurves of V10, the RR Lyrae, and
V13 and V14, the two unclassified systems, from the LCO
dataset. Figure \ref{wuma6} contains the lightcurve of V6, the only
member of NGC 3201 among our set of variable stars. V11, the Algol
system, and V12, the detached system, are presented in Figures
\ref{algol} and \ref{detached}, respectively. In all of these three
Figures, $V$ data are in the bottom panel, $I$ data in the top
panel. The error bars represent the photometric errors associated with
that particular measurement of the star's magnitude. No reddening
correction was applied to the lightcurves.

\subsection{Rucinski Magnitudes}

W Ursa Majoris binaries are systems in which the two components are in
physical contact with their Roche equipotential lobes and at their
inner Langrangian Point (see \citet{mateo93}, \citet{rucinski85a}, and
\citet{rucinski85b} for more detailed descriptions of W UMa
systems). Due to the good thermal contact between the two components,
the two stars are assumed to have the same surface temperature. The
total luminosity of the system can therefore be defined as
\citep{rucinski1995,mateo1996}

\begin{equation}
L=KT^{4}S(q,P,f)M_{p}^{\frac{2}{3}}(1+q)^{\frac{2}{3}}P^{\frac{4}{3}},
\end{equation}

where $K$ consists of well-known constants, $T$ is the common surface
temperature, $M_{p}$ and $M_{s}$ are the masses of the primary and
secondary component, respectively, $q = M_{s} / M_{p}$, and $P$ is the
orbital period of the system. $S$ represents the total stellar surface
area as defined by the Roche geometry, dynamical properties of the
binary, and the fill-out factor $f$ by which the stellar surfaces
extend beyond the inner critical Roche surface.

\citet{rucinski1994,rucinski1995,rucinski2000} converted
the above equation into the more convenient form

\begin{equation}
M_{V}^{color}=a_{0}+a_{1}\log P+a_{2} color,
\end{equation}

where $M_{V}$ is the absolute $V$ magnitude, $P$ is the period
in days, and the color is reddening-free. For $V-I$ color, this equation
is \citep{rucinski2000}

\begin{equation}
M_{V}^{VI}=-4.43\log P+3.63 (V-I)_{0}-0.31; \sigma \sim 0.29.
\end{equation}

Table 2 shows the thus calculated absolute magnitudes and distance
moduli for the W UMa binaries in the field of NGC 3201. The distance
to NGC 3201 was calculated in BM01 to be $4.5 \pm 0.03$ kpc; the value
found in \citet{harris1996} is 5.2 kpc. The corresponding distance
moduli are $V_{0}-M_{V}=13.58$ and 13.26, respectively.\footnote{These
values represent the apparent distance moduli corrected for
extinction.} We note that the absolute magnitudes and distance moduli
in Table 2 were calculated under the assumption that the W UMa system
under investigation is suffering the full extinction between us and
the cluster. Since the spectroscopy results indicated that no W UMa
system except V6 is associated with NGC 3201 (see following
Subsection), and the Rucinski magnitudes for some of the variables
indicate that they are foreground stars, this assumption might be
incorrect in some cases. That is, for some of the non-members, the
color might not be the correct, reddening-free value.

The distance modulus of the GC member V6 calculated in Table
2 is 13.873, larger than the NGC 3201 distance moduli in both BM01 (by
$\sim 2\sigma$) and \citet{harris1996} (by $\sim 1\sigma$). One way to
{}``force{}'' V6 to be at the correct distance would be to change the
reddening zero point to the cluster and thus change the variable's
intrinsic color. Since this correction would force the distance to go
down, the reddening zero point obtained this way would decrease to
$E_{V-I} \sim 0.02$, significantly below the BM01 value of 0.15. At
this point, we can therefore not offer any conclusive reason why V6's
distance modulus is slightly too high, other than the possibility
that this one particular binary would represent an outlier (by about
1-2$\sigma$) in Rucinski's empirically determined relation.

\subsection{Cluster Membership}

A first indication of whether an eclipsing binary star system in the
field of a globular cluster is associated with that cluster is, of
course, its location with respect to the cluster center (we note that
we cannot study the very center of the cluster due to crowding). The
tidal radius of NGC 3201 is 28.45 arcmin \citep{harris1996}. Thus, all
the binary systems we find are well within the tidal radius. A more
powerful membership criterion is a binary's location in the CMD. Based
on the CMD of NGC 3201 (see Figure \ref{cmd_vars}), V5, V7, V8, V10,
V13 and V14 are not associated with the cluster. Furthermore, based on
their distance moduli, the W UMa systems V3, V4, V5, and V8 seem to be
foreground stars while V9 seems to be a background star.  V1, V2, and V6,
however, seem promising cluster member candidates within error
estimates, with the only difference being that the reddening
calculated for V1 and V2 is significantly lower than the one for V6
(see Table 2).

The final verdict on membership, however, can only be provided by
spectroscopic observations. NGC 3201 has a distinct systemic
(retrograde) velocity of approximately 500 km/s
\citep{harris1996,cote1994}, so a single spectrum of a binary system
with sufficient resolution and signal-to-noise ratio can be used to
establish cluster membership.

During the nights of March 2 - 5, 2001, we performed spectroscopic
follow-up observations of our eclipsing binary star candidates at the
CTIO 4m Telescope with Arcon and the RC Spectrograph. Our setup
included a 3k x 1k Loral CCD and the KPGLG grating (860 l/mm; 1st
order blaze = 11000 \AA) in second order. The wavelength coverage
extended from 3800 to 5100 \AA, with a resolution of about 0.4
\AA/pixel.  Our observing targets were V1, V2, V3, V4, V6, V11 and
V12. V9, a potential cluster member given its location in the CMD, was
too faint to observe with our setup at CTIO.

In order to check the cluster membership of V9, we obtained three
spectra during the night of March 19, 2001, using the Magellan 1
Telescope with the Boller \& Chivens Spectrograph and a setup similar
to the one at the CTIO 4m Telescope but with lower resolution
(approximately 1.4 \AA/pix). Our wavelength coverage extended from
3900 to 5300 \AA.

To our great disappointment during the first night of the CTIO spectroscopy
run, one spectrum after the next eliminated our candidates from cluster
membership consideration. The only system with a systemic velocity
equal to that of NGC 3201 is V6, a blue straggler candidate (see Figure
\ref{cmd_vars}) with a very small amplitude (see Figure \ref{wuma6}). 

Our spectroscopy results are summarized in Table 3. The systemic
velocities of the binaries were determined using the IRAF FXCOR
package, using radial velocity standards as templates. We estimate our
error in the velocity determinations of the candidates to be around 20
km/s for the CTIO data (except V6), based on measurements performed on
the velocity standards we used. Due to the lower resolution of the
spectrum we obtained at Magellan, we estimate V9's velocity
uncertainty to be approximately 50 km/s. The difference in spectral
types between V6 and our radial velocity standards made it impossible
to determine its velocity using FXCOR. Instead, we calculated its
Doppler velocity from line profile fitting using IRAF's SPLOT
package. The error quoted for that star is simply the rms of $\sim 10$
individual line measurements from different spectra of the star.

We further note that for all of our variables, except for our CTIO
photometry targets (V7-9 as well as V11 and V12), we do not have phase
information obtained more recently than three years prior to the
spectroscopy run. The propagation of the period determination error
(see Tables 1 and 2) therefore makes it impossible to constrain the
precise phasing of most of our candidates. The uncertainty in the
calculation of the systemic velocities of the binary candidates due to
the radial motion of the two components is not accounted for in our
estimates.\footnote{ A binary's period may, of course, also be
calculated from the radial velocity curve. Since we were able to
eliminate our observing targets (except V6) from cluster membership
consideration after obtaining only a few spectra for each system, we
refrained from re-observing them and can therefore not determine the
phase of the binary at the time of observation.  }

Since NGC 3201's systemic velocity is so high (500 km/s), cluster
membership based on the comparison between systemic velocity of the
variable and the cluster may be determined without phasing
information.  The velocity amplitudes of the components of a binary
system relate to the masses and the period of the system as follows
\citep{paczynski96}:

\begin{equation}
\frac{K_{1}+K_{2}}{\sin i}=386.4 km s^{-1}\left(
\frac{M_{1}+M_{2}}{3 M_{solar}}\frac{0.5 days}{P_{orb}}\right)
^{\frac{1}{3}}.
\end{equation}

If one assumes that the orbits of the W UMa systems V1-4 and V9 are
circular (which is evidenced by the fact that the minima of the
lightcurves are separated by half a period), that \( \sin i=1 \) (the
system is seen edge-on), and that the masses and thus velocity
amplitudes of the two components are equal, one needs only the sum of
the masses and the orbital period to get an estimate of the velocity
amplitudes.  From the isochrones of D. VandenBerg 2000 (private
communication; based on evolutionary models by \citet{vandenberg00})
for a cluster age of anywhere inbetween 14 and 18 Gyrs and
{[}Fe/H{]}=-1.41, one obtains masses for main-sequence stars of
$(V-I)_{0} \sim 0.7$ of approximately 0.7 $M_{solar}$ (for each
component). Finally, given the average period of these systems of
$\sim 0.36$ days, we estimate the typical velocity amplitude to be
$K_{j} \sim 165 km s^{-1}$. We note, of course, that non-member
binaries might have a different metallicity and thus mass for a given
color, but since $K \propto M^{1/3}$, this difference would not
significantly alter this rough estimate. A deviation from an edge-on
configuration would decrease the calculated velocity amplitudes, i.e.,
the value given above may be regarded as an upper limit.

For V11 and V12, the masses of the individual components are probably
slightly larger (bluer $V-I$ color), but the periods are much larger,
especially in the case of V12. Retaining the same assumptions as for
the W UMa systems, one obtains velocity amplitudes closer to 150 km/s
for V11, and 80 km/s for V12.

In order to for a binary system to be a member of NGC 3201, its
systemic velocity would have to fall within (165 km/s + \( \sigma \))
of 500 km/s (NGC 3201's velocity). As one may easily see in Table 3,
only V6 survives this criterion.

We obtained approximately 25 spectra of V6 at various phase angles. 
These results, combined with our photometry data, may be used to 
calculate the component masses of the binary BS system. The analysis 
of the radial velocity curve and subsequent calculation of the stellar 
masses will be addressed in a future publication.

\section{Summary and Concluding Remarks}

The low-latitude GC NGC 3201 was probed for the existence of
photometrically variable stars in the magnitude range of approximately
\( 16.5<V<20 \) and for periods between roughly 0.2 and 5 days. We
detected 14 variable star candidates in the field of which most are
eclipsing binaries of the W UMa type. Our spectroscopic follow-up
observations revealed that only one of the variables, a BS W UMa EB,
is associated with the cluster itself. Due to the low-latitude
location of the GC, the high contamination of Galactic disk stars in
the field of NGC 3201 seems to have manifest itself 13 times in our
analysis.

Our confirmation of V6 as a member increases the number of known
binary BSs in GCs. \citet{mateo90} predicted that 3\%-15\% of all BSs
in GCs should be photometrically observable binaries. In the case of
NGC 3201 there are nine total BSs \citep{sarajedini93}. Our detection
of one BS member binary is therefore consistent with that prediction.
Thus, the membership of V6 supports the basic \citet{mateo90} model
that BSs evolve from primordial or, at least, long-lived binaries
which coalesce after a lifetime of slow angular momentum loss.

\begin{figure}
{\centering
\resizebox*{0.6\textwidth}{0.4\textheight}{\includegraphics{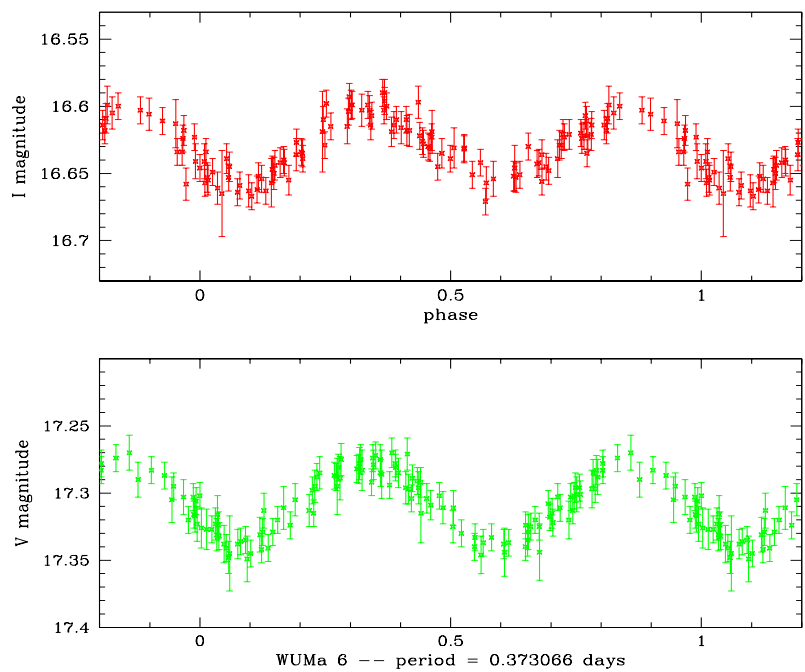}}
\par}
\caption{\label{wuma6}The lightcurve of V6, a blue straggler candidate and
the only variable star in our sample which is a member of NGC 3201.
No extinction correction is applied.}
\end{figure}

\begin{figure}
{\centering
\resizebox*{0.6\textwidth}{0.4\textheight}{\includegraphics{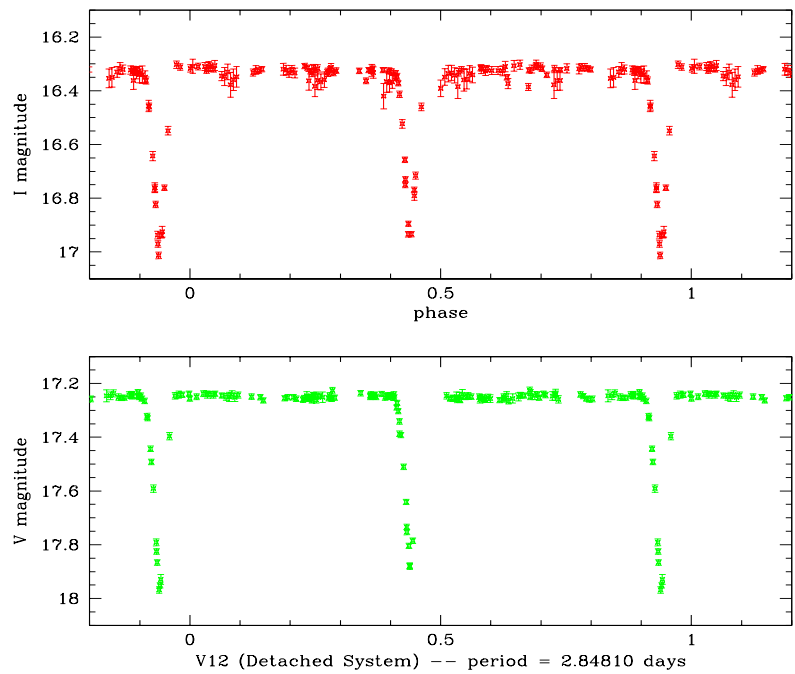}}
\par}
\caption{\label{detached}The lightcurve of V12, a detached binary star
system.  The data for this lightcurve were taken both at LCO and CTIO,
and they are not corrected for extinction. V12 is not associated with
NGC 3201. }
\end{figure}

Despite the fact that we did not identify a large number of NGC 3201
member binaries, we have nevertheless shown the potential to do just
that for the other globular clusters in our survey (provided there are
binaries in those clusters). The fact that we successfully identified
a BS EB with an $V$ amplitude of around 0.07 mag (see Figure
\ref{wuma6}), a detached system with duty cycle of around 0.1 (see
Figure \ref{detached}), and that we determined periods to a precision
of seconds or less (see Section 3.3) gives us confidence that we have
the dataset and the methods to detect a number of EBs in our target
GCs. Using the combination of our 1m telescope photometry dataset and
the capabilities of Magellan 1 for the spectroscopy follow-up
observations, we should be able to determine the cluster membership
for binaries in the southern and equatorial GCs of our sample in the
near future.

\acknowledgements{}

This research was funded in part by NSF grants AST 96-19632 and
98-20608. We would like to thank A. Udalski for his help with the AoV
algorithm. We would also like to express our most sincere gratitude to
the support staff and night assistants at CTIO, Magellan, and
especially LCO without whose help and determination these observations
would not have been able to produce the results presented
here. Finally, we thank the anonymous referee for his/her thorough
analysis of the manuscript and insightful comments and suggestions.

\clearpage


\begin{deluxetable}{cccclcc}
\tabletypesize{\scriptsize}
\tablecaption{Variable Stars in the Field of NGC 3201}
\tablewidth{0pt}
\tablehead{
\colhead{Var. No.} &
\colhead{type} &
\colhead{RA (2000)} &
\colhead{Dec (2000)} &
\colhead{period (days)} &
\colhead{$V_{bright}$} &
\colhead{$I_{bright}$}
} 
\startdata
V1&
W UMa&
10:16:36.92&
-46:22:29.3&
0.303587(28)&
18.054(17)&
17.258(20)\\
V2&
W UMa&
10:17:07.73&
-46:30:18.2&
0.345095(42)&
18.237(14)&
17.319(19)\\
V3&
W UMa&
10:17:13.75&
-46:27:54.7&
0.377114(43)&
17.189(21)&
16.352(21)\\
V4&
W UMa&
10:17:17.18&
-46:27:37.5&
0.44179(55){*}&
16.850(17)&
15.965(22)\\
V5&
W UMa&
10:17:52.93&
-46:34:06.7&
0.276216(31)&
19.847(21)&
18.380(32)\\
V6&
W UMa&
10:17:59.08&
-46:33:25.7&
0.37307(39){*}&
17.270(13)&
16.599(19)\\
V7&
W UMa&
10:18:56.03&
-46:36:10.4&
1.0800(90)&
18.658(15)&
17.064(19)\\
V8&
W UMa&
10:18:46.01&
-46:30:13.8&
0.30642(75)&
18.824(14)&
17.577(19)\\
V9&
W UMa&
10:18:31.97&
-46:37:32.9&
0.33248(61)&
19.609(22)&
18.483(27)\\
V10&
RR Lyrae&
10:18:03.86&
-46:17:48.7&
0.592920(53)&
17.088(12)&
16.446(22)\\
V11&
Algol&
10:18:32.74&
-46:30:42.5&
0.702127(99)&
17.728(13)&
16.932(19)\\
V12&
Detached&
10:17:25.72&
-46:20:12.4&
2.84810(98)&
17.225(14)&
16.301(22)\\
V13&
Unclass. 1&
10:17:04.83&
-46:26:39.7&
1.2080(44){*}&
20.203(35)&
18.529(29)\\
V14&
Unclass. 2&
10:18:36.28&
-46:22:06.2&
2.160(13){*}&
18.915(16)&
17.108(21)\\
\enddata

\tablecomments{
\begin{itemize}
\item Errors in parentheses indicate the uncertainty in last two digits,
i.e., $0.303587(28) = 0.303587 \pm 0.000028$ and $2.160(13) = 2.160 \pm 0.013.$
\item Photometry errors are the result of adding in quadrature the DoPHOT
photometry error for the instrumental magnitude and the rms of the
standard star solution (see BM01). 
\item The error in the period corresponds to the full width at half maximum
(fwhm) of the peak in the AoV power spectrum corresponding to the
correct frequency. For the determination of this error, only $V$
data were used, except for the cases of V7, V8 and V13 where we used
$I$ data. In some instances, the peak in the power spectrum
corresponding to the correct period was assigned essentially the same
power as directly neighboring peaks (i.e., it would be hard to pick
{}``by eye{}'' which one is the correct one). In these cases, we
estimated the period error to be the distance between the two neighboring
peaks (with the peak corresponding to the correct period in the middle).
These cases are marked by an asterisk. 
\end{itemize}
}

\end{deluxetable}

\clearpage


\begin{deluxetable}{clcccc}
\tabletypesize{\scriptsize}
\tablecaption{Rucinski Magnitudes and Distance Moduli for W UMa-type Binaries}
\tablewidth{0pt}
\tablehead{
\colhead {system} &
\colhead {period (days)\tablenotemark{a}} &
\colhead {$E_{V-I}$\tablenotemark{b}} &
\colhead {$(V-I)_{0, bright}$\tablenotemark{c}} &
\colhead {$M_{V(Rucinski)}$} &
\colhead {$V_{0}-M_{V}$}
}
\startdata
V1\tablenotemark{d}&
0.303587(28)&
0.131(55)&
0.665&
4.397&
13.405\\
V2&
0.345095(42)&
0.202(34)&
0.716&
4.336&
13.513\\
V3&
0.377114(43)&
0.183(38)&
0.654&
3.940&
12.898\\
V4&
0.44179(55)&
0.183(38)&
0.702&
3.810&
12.689\\
V5&
0.276216(31)&
0.305(65)&
1.162&
6.383&
12.878\\
V6&
0.37307(39)&
0.366(42)&
0.305&
2.694&
13.873\\
V7&
1.0800(90)&
0.352(81)&
1.242&
4.050&
13.932\\
V8&
0.30642(75)&
0.275(61)&
0.972&
5.494&
12.802\\
V9&
0.33248(61)&
0.368(67)&
0.758&
4.560&
14.343\\
\enddata

\tablenotetext{a}{As in Table 1, errors in parentheses indicate the
uncertainty in last two digits. The period errors are the same as in
Table 1.}  
\tablenotetext{b}{$E_{V-I}$ contains the reddening zero
point, calculated in BM01 to be 0.15 mag. As mentioned above, this
value assumes that the binary suffers the full extinction along the
line of sight to the GC (which is not necessarily correct if the
binary is not a cluster member).  The errors for $E_{V-I}$ are the
random errors in the determination of the \textit{differential
reddening.} That is, a possible systematic error in the determination
of the reddening zero point (described in BM01) is not included in
this estimate.}  
\tablenotetext{c}{$(V-I)_{0, bright}$ is the
dereddened color at maximum light.}
\tablenotetext{d}{V1 is located in a region whose differential
reddening along the line of sight is lower than the one of the
fiducial region whose overall $E_{V-I}$ is 0.15 mag. Its total
$E_{V-I}$ is therefore calculated to be lower than the reddening zero
point.}

\tablecomments{ Rucinski quotes the scatter in his relation
\citep{rucinski1994,rucinski1995} to be 0.29 mag in the calculation of
the absolute $V$ magnitudes (corresponding to an uncertainty of
approximately 13\% in distance). Since this uncertainty is far larger
than the quadratic sum of all our random errors, we refrain from a
detailed error analysis for the Rucinski magnitudes and distance moduli.  }

\end{deluxetable}

\clearpage


\begin{deluxetable}{ccc}
\tabletypesize{\scriptsize}
\tablecaption{Systemic Velocities of Member Candidates}
\tablewidth{0pt}
\tablehead{
\colhead{variable} &
\colhead{systemic velocity (km/s)} &
\colhead{error (km/s)}} 
\startdata
V1&
12&
20\\
V2&
6&
20\\
V3&
9&
20\\
V4&
20&
20\\
V6&
513&
44\\
V9&
24&
50\\
V11&
1&
20\\
V12&
0&
20\\
\enddata

\tablecomments{
\begin{itemize}
\item The variables not listed in this table but listed in Tables 1
and 2 were considered to be non-members based on their location in the
CMD.
\item For an explanation on how velocities and errors were determined,
see text.
\item The systemic velocity of NGC 3201 is $\sim 500 km/s$ \citep{harris1996,
cote1994}.
\end{itemize}
}

\end{deluxetable}

\clearpage


\clearpage

\clearpage

\clearpage

\clearpage

\clearpage

\clearpage

\clearpage

\end{document}